\newcommand{\extended}[1]{}    

\documentclass{article}

\usepackage{graphicx}
\usepackage{mathrsfs}
\usepackage[T1]{fontenc}
\usepackage{tikz}
\usetikzlibrary{decorations,decorations.pathmorphing,arrows,shapes,%
  automata,backgrounds,fit,calc,petri,patterns,matrix}
\usepackage{algorithmic}
\usepackage{verbatim}
\usepackage{color}
\usepackage{stmaryrd}
\usepackage{enumerate}
\usepackage{xspace}
\usepackage{balance} 
\usepackage{multirow}
\usepackage{adjustbox}
\usepackage{hyperref}
\usepackage{authblk}

\usepackage{wojtek15logics,wojtek15other}

\newcommand{\BibTeX}{\rm B\kern-.05em{\sc i\kern-.025em b}\kern-.08em\TeX}
\pgfarrowsdeclare{arrow30}{arrow30}
{
  \arrowsize=0.3pt
  \advance\arrowsize by.275\pgflinewidth%
  \pgfarrowsleftextend{+-\arrowsize}
  \advance\arrowsize by.5\pgflinewidth
  \pgfarrowsrightextend{+\arrowsize}
}
{
  \arrowsize=0.3pt
  \advance\arrowsize by.275\pgflinewidth%
  \pgfsetdash{}{+0pt}
  \pgfsetroundjoin
  \pgfpathmoveto{\pgfqpoint{1\arrowsize}{0\arrowsize}}
  \pgfpathlineto{\pgfqpoint{-12\arrowsize}{2.5\arrowsize}}
  \pgfpathlineto{\pgfqpoint{-12\arrowsize}{-2.5\arrowsize}}
  \pgfpathclose
  \pgfusepathqfillstroke
}

\definecolor{tucgreen}{RGB}{0,140,79}

\newcommand{\STVAGR}{\textbf{STV+AGR}\xspace}
\newcommand{\STV}{\textbf{STV}\xspace}

\title{STV+AGR: Towards Practical Verification of Strategic Ability Using Assume-Guarantee Reasoning}

\begin{document}

\author[1,2]{Damian Kurpiewski}
\author[2,1]{{\L}ukasz Mikulski}
\author[1,3]{Wojciech Jamroga}

\affil[1]{Institute of Computer Science, Polish Academy of Sciences, Warsaw, Poland}
\affil[2]{Faculty of Mathematics and Computer Science, Nicolaus Copernicus University, Toru{\'n}, Poland}
\affil[3]{Interdisciplinary Centre for Security, Reliability and Trust, SnT, University of Luxembourg, Luxembourg}

\date{}
\setcounter{Maxaffil}{0}
\renewcommand\Affilfont{\itshape\small}

\maketitle

\begin{abstract}
We present a substantially expanded version of our tool \textbf{STV} for strategy synthesis and verification of strategic abilities.
The new version provides a web interface and support for assume-guarantee verification of multi-agent systems.
\end{abstract}







\section{Introduction}\label{sec:intro}

Model checking of multi-agent systems (MAS) allows for formal (and, ideally, automated) verification of their relevant properties.
Algorithms and tools for model checking of \emph{strategic abilities}~\cite{Alur02ATL,Schobbens04ATL,Chatterjee10strategylogic,Mogavero14behavioral} have been in development for over 20 years~\cite{Alur98mocha-cav,Chen13prismgames,Busard14improving,Huang14symbolic-epist,Cermak14mcheckSL,Lomuscio17mcmas,Cermak15mcmas-sl-one-goal,Belardinelli17publicActions,Belardinelli17broadcasting,Jamroga19fixpApprox-aij,Kurpiewski21stv-demo}.
Unfortunately, the problem is hard, especially in the realistic case of agents with imperfect information~\cite{Schobbens04ATL,Bulling10verification,Dima11undecidable}.

In this paper, we propose a new extension of our experimental tool \textbf{STV}~\cite{Kurpiewski19stv-demo,Kurpiewski21stv-demo} that facilitates compositional model checking of strategic properties in asynchronous MAS through assume-guarantee reasoning (AGR)~\cite{Pnueli84assGuar,Clarke89assGuar}.
The extension is based on the preliminary results in~\cite{Mik22assGua}, itself an adaptation of the AGR framework for liveness specifications from~\cite{Lomuscio10assGar, Lomuscio13assGar}.

\section{Application Domain}\label{sec:domain}

Many important properties of MAS refer to \emph{strategic abilities} of agents and teams.
For example, the \ATLs formula $\coop{taxi}\Always\neg\prop{fatality}$ says that the autonomous cab can drive in such a way that no one gets ever killed, and $\coop{taxi,passg}\Sometm\,\prop{destination}$ expresses that the cab and the passenger have a joint strategy to arrive at the destination, no matter what the other agents do.
Another intuitive set of strategic requirements is provided by properties of secure voting systems~\cite{Ryan10atemyvote,Tabatabaei16expressing}.
As shown by case studies~\cite{Jamroga18Selene,Jamroga20Pret-Uppaal,Jamroga21natstrat-voting} practical verification of such properties is still infeasible due to state-space and strategy-space explosion.
\textbf{STV+AGR} addresses the specification and verification of such properties, as well as a user-friendly creation of models to be verified.

\section{Simple Voting Scenario}\label{sec:scenarios}

\begin{figure}[t]\centering
\begin{tabular}{@{}c@{}}
\begin{tikzpicture}[->,auto,>=arrow30,node distance=1.5cm,font=\tiny]
 \node[initial, initial where=above,state] (A)                      {$?,?$};
\node[state]         (B1) [below left of=A]   {$T,?$};
\node[state]         (B2) [below right of=A]  {$F,?$};
\node[state]         (B3) [left of=B1]        {$?,T$};
\node[state]         (B4) [right of=B2]       {$?,F$};
\node[state]         (C1) [below of=B1]       {$F,T$};
\node[state]         (C2) [below of=B2]       {$T,F$};
\node[state]         (C3) [below of=B3]       {$T,T$};
\node[state]         (C4) [below of=B4]       {$F,F$};

\path (A) edge [loop below] node {$?\times ?$} (A); 

\path (A) edge [bend right] node [swap] {$\forall\times \lnot ?$} (B3); 
\path (A) edge [bend left] node {$\forall\times \lnot ?$} (B4); 
\path (A) edge node [swap] {$\lnot ?\times\forall$} (B1); 
\path (A) edge node {$\lnot ?\times\forall$} (B2); 

\path (B1) edge [loop left] node [below] {$\forall\times ?$} (B1); 
\path (B3) edge [loop left] node [below] {$?\times \forall$} (B3); 
\path (B2) edge [loop right] node [below] {$\forall\times ?$} (B2); 
\path (B4) edge [loop right] node [below] {$?\times \forall$} (B4); 

\path (B1) edge node [rotate = 45, xshift=-1] {$\forall\times \lnot ?$} (C3); 
\path (B1) edge node [swap, rotate = -35] {$\forall\times \lnot ?$} (C2); 

\path (B2) edge node [swap, rotate = -45, xshift=1] {$\forall\times \lnot ?$} (C4); 
\path (B2) edge node [rotate = 35] {$\forall\times \lnot ?$} (C1); 

\path (B3) edge node [swap, rotate = -45, xshift=1] {$\lnot ? \times\forall$} (C1); 
\path (B3) edge node [swap] {$\lnot ? \times\forall$} (C3); 

\path (B4) edge node [rotate = 45, xshift=-1] {$\lnot ? \times\forall$} (C2); 
\path (B4) edge node {$\lnot ? \times\forall$} (C4); 

\path (C1) edge [loop left] node [below] {$\forall\times\forall$} (C1); 
\path (C2) edge [loop right] node [below] {$\forall\times\forall$} (C2); 
\path (C3) edge [loop left] node [below] {$\forall\times\forall$} (C3); 
\path (C4) edge [loop right] node [below] {$\forall\times\forall$} (C4); 
\end{tikzpicture}
\end{tabular}

\bigskip
\begin{tabular}{@{}c@{}}
\begin{tikzpicture}[->,auto,>=arrow30,node distance=1.5cm,font=\tiny]
 \input{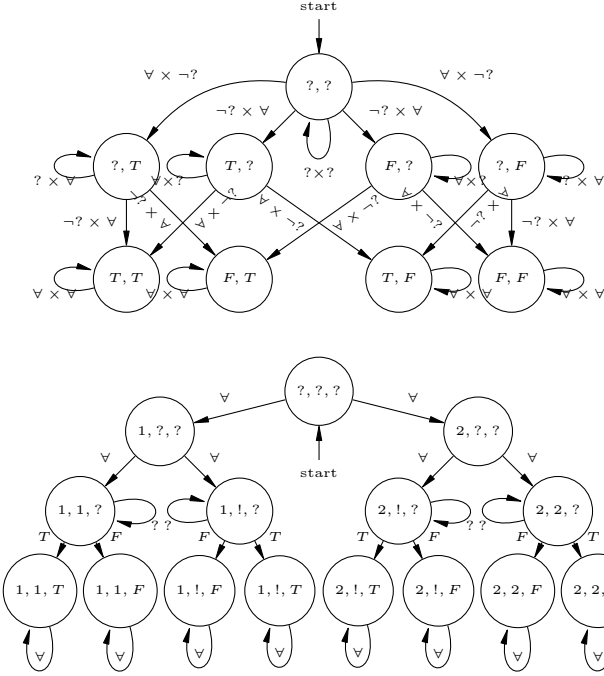}
\end{tikzpicture}
\end{tabular}
\caption{Two modules: a coercer[2] (up) and a voter (down)}
\label{fig:coercer}
\label{fig:voter}
\end{figure}


To present the capabilities of \textbf{STV+AGR}, we designed an asynchronous version of the Simple Voting scenario~\cite{Jamroga19fixpApprox-aij}.
The model consists of two types of agents, presented in Figure~\ref{fig:voter}, and described below.

\para{Voter.}
Every voter agent has three local variables:
\begin{itemize2}
    \item $vote_i$: the vote being cast ($?, 1$, or $2$);
    \item $reported_i$: the vote value presented to the coercer ($?, 1, 2$ or $!$), where $!$ means that the voter decided not to share her vote with the coercer;
    \item $pstatus_i$: the punishment status ($?, T$ or $F$).
\end{itemize2}
Each voter $i$ can also see the value of the $pun_i$ variable of the coercer.

The voter first casts her vote, then decides whether to share its value with the coercer.
Finally, she waits for the coercer's decision to punish her or to refrain from punishment.

\para{Coercer.}
The coercer[k] has one local variable for each of $k$ voters:
\begin{itemize2}
    \item $pun_i$: whether the voter $i$ was punished or not ($?, T$ or $F$).
\end{itemize2}
Moreover, he can observe the value of $reported_i$ for each voter $i$.

The coercer has two available actions per voter: to punish the voter or to refrain from punishment.

\section{Formal Background}\label{sec:background}

\para{Modules.} The main part of the input is given by a set of asynchronous modules inspired by~\cite{Lomuscio13assGar},
where local states are labelled with valuations of state variables.
The transitions are valuations of input variables controlled by the other modules.
The multi-agent system is defined by a composition of its modules.

\para{Strategies.} A strategy is a conditional plan that specifies what the agent(s) are going to do in every possible situation.
Here, we consider the case of \emph{imperfect information memoryless strategies}, represented by functions from the agent's local states (or, equivalently, its epistemic indistinguishability classes) to its available actions.
The \emph{outcome} of a strategy from state $q$ consists of all the infinite paths starting from $q$ and consistent with the strategy.

\para{Logic.} Given a model $M$ and a state $q$ in the model, the $\oneATLsX$~\cite{JPSDM20} formula $\coop{a}\varphi$ holds in the pointed model $(M,q)$ iff there exists a strategy for agent $a$ that makes $\varphi$ true on all the outcome paths starting from any state indistinguishable from $q$.
The semantics of coalitional abilities is analogous, for joint strategies of coalitions.

\para{Assume-guarantee reasoning.} The main idea is to cope with the state-space explosion
by decomposing the goal $\varphi$ of coalition $C$ into local goals $\varphi_i, i\in C$, and
verify them one by one against abstractions of each agent's environment.
An abstraction for $i$ is obtained by defining a single module, called the \emph{assumption},
which \emph{guarantees} that all the paths present in the original system have their counterparts in the composition of module $i$ and its associated assumption.
Moreover, we use a distance between modules, based on shared synchronization actions, so that only ``close'' agents are taken into account when preparing the assumption for $i$.
This way, one can deduce the existence of a joint strategy to obtain $\varphi$
from the existence of individual strategies that achieve local goals $\varphi_i$.

\para{Automated generation of assumptions.} The main difficulty in using assume-guarantee reasoning is how to define the right assumptions for the relevant modules.
To this end, we propose an automated procedure that generates the assumptions, based on the subset of modules that are ``close'' the given module $M_i$.
The abstraction is obtained by composing all the ``close'' modules, abstracting away their state labels and variables except for the ones that are input to $M_i$, as well as removing all their input variables which are not state variables of $M_i$.

\section{Technology}\label{sec:technology}

\STVAGR does \emph{explicit-state model checking}.
That is, the global states and transitions of the model are represented explicitly in the memory of the verification process.
The tool includes the following new functionalities.

\para{User-defined input.}
The user can load and parse the input specification from a text file that defines the groups of modules.
The modules are local automata representing the agents.
The groups define the partition for the assume-guarantee verification.
Each group that describes the part of the coalition must also define the formula to be verified.

\para{Web-based graphical interface.}
The generated models and the verification results are visualised in the intuitive web-based graphical interface.
The GUI is implemented in Typescript and uses the Angular framework.

\begin{table}[t]
    \begin{adjustbox}{width=1.05\columnwidth,center}
    \begin{tabular}{|c|c|c|c|c|c|c|c|c|}
    \hline
    \multirow{2}{*}{\textbf{V}}  & \multicolumn{4}{c|}{\textbf{Monolithic model checking}}       & \multicolumn{4}{c|}{\textbf{Assume-guarantee verification}}   \\ \cline{2-9}
                                    & \textbf{\#st} & \textbf{\#tr} & \textbf{DFS} & \textbf{Apprx} & \textbf{\#st} & \textbf{\#tr} & \textbf{DFS} & \textbf{Apprx} \\ \hline
    2                           & 529           & 2216          & <0.1         & <0.1/Yes           & 161           & 528          & <0.1         & <0.1/Yes       \\ \hline
    3                           & 12167         & 127558          & <0.1         & 0.8/Yes          & 1127          & 7830          & <0.1         & <0.1/Yes        \\ \hline
    4                           & 2.79e5        & 6.73e6 &\multicolumn{2}{c|}{memout}               & 7889         & 1.08e5         & <0.1         & 0.8/Yes           \\ \hline
    5                           & \multicolumn{4}{c|}{memout}                                       & 5.52e4         & 1.45e6         & <0.1         & 11/Yes          \\ \hline
    \end{tabular}
    \end{adjustbox}
\vspace{0.2cm}
\caption{Results of assume-guarantee verification the asynchronous variant of Simple Voting (times given in seconds)}
\vspace{-0.6cm}
\label{tab:res}
\end{table}

\para{Evaluation.}
The assumption-guarantee scheme has been evaluated on the asynchronous variant of Simple Voting, using formula $\varphi \equiv \coop{\mathit{Voter}_1}\Always(\lnot \prop{pstatus_1} \lor \prop{voted_1}=1)$.
Note that the coalition consisted of only one agent, which made the decomposition of the formula trivial.
The results are presented in Table~\ref{tab:res}.
The first column describes the configuration of the benchmark, i.e., the number of voters.
Then, we report the performance of model checking algorithms that operate on the explicit model of the whole system vs.~assume-guarantee verification.
\emph{DFS} is a straightforward implementation of depth-first strategy synthesis. \emph{Apprx} refers to the method of fixpoint-approximation~\cite{Jamroga19fixpApprox-aij}; besides the time, we also report if the approximation was conclusive.

\section{Usage}\label{sec:usage}

The tool is available at \href{http://stv.cs-htiew.com}{stv.cs-htiew.com}.
The video demonstration of the tool is available at \href{https://youtu.be/1DrmSRK1fBA}{youtu.be/1DrmSRK1fBA}.
Example specifications can be found at \href{http://stv-docs.cs-htiew.com}{stv-docs.cs-htiew.com}.
The current version of \textbf{STV+AGR} allows to:
\begin{itemize}
\item Generate and display the composition of a set of modules into the model of a multi-agent system;
\item Generate and display the automatic assumption, given a module and a distance bound;
\item Provide local specifications for modules, and compute the global specification as their conjunction;
\item Verify a $\oneATLsX$ formula for a given system (using the verification methods available in the \STV package);
\item Verify a $\oneATLsX$ formula for a composition of a module and its automatic assumption (using the methods in \STV);
\item Verify a $\oneATLsX$ formula for a composition of a module and a user-defined assumption (using the methods in \STV);
\item Display the verification result, including the relevant truth values and the winning strategy (if one exists).
\end{itemize}

\section{Conclusions}\label{sec:conlusions}

Much complexity of model checking for strategic abilities is due to the size of the model.
\STVAGR addresses the challenge by implementing a compositional model checking scheme, called assume-guarantee verification. 
No less importantly, our tool supports user-friendly modelling of MAS, and automated generation of abstractions that are used as assumptions in the scheme.

\section*{Acknowledgement}
  The authors thank Witold Pazderski and Yan Kim for assistance with the web interface.
  The work was supported by NCBR Poland and FNR Luxembourg under the PolLux/FNR-CORE project STV (POLLUX-VII/1/2019), as well as the CHIST-ERA grant CHIST-ERA-19-XAI-010 by NCN Poland (2020/02/Y/ST6/00064).

\bibliographystyle{plain}
\bibliography{biblio}

\begin{thebibliography}{10}

\bibitem{Alur02ATL}
R.~Alur, T.A. Henzinger, and O.~Kupferman.
\newblock {A}lternating-time {T}emporal {L}ogic.
\newblock {\em J. of the ACM}, 49:672--713, 2002.

\bibitem{Alur98mocha-cav}
R.~Alur, T.A. Henzinger, F.Y.C. Mang, S.~Qadeer, S.~Rajamani, and S.~Tasiran.
\newblock {MOCHA}: Modularity in model checking.
\newblock In {\em Proc. of {CAV}'98}, volume 1427 of {\em LNCS}, pages
  521--525. Springer, 1998.

\bibitem{Belardinelli17broadcasting}
F.~Belardinelli, A.~Lomuscio, A.~Murano, and S.~Rubin.
\newblock Verification of broadcasting multi-agent systems against an epistemic
  strategy logic.
\newblock In {\em Proc. of {IJCAI}'17}, pages 91--97, 2017.

\bibitem{Belardinelli17publicActions}
F.~Belardinelli, A.~Lomuscio, A.~Murano, and S.~Rubin.
\newblock Verification of multi-agent systems with imperfect information and
  public actions.
\newblock In {\em Proc. of {AAMAS}'17}, pages 1268--1276, 2017.

\bibitem{Bulling10verification}
N.~Bulling, J.~Dix, and W.~Jamroga.
\newblock Model checking logics of strategic ability: Complexity.
\newblock In {\em Specification and Verification of Multi-Agent Systems}, pages
  125--159. Springer, 2010.

\bibitem{Busard14improving}
S.~Busard, C.~Pecheur, H.~Qu, and F.~Raimondi.
\newblock Improving the model checking of strategies under partial
  observability and fairness constraints.
\newblock In {\em Formal Methods and Software Engineering}, volume 8829 of {\em
  LNCS}, pages 27--42. Springer, 2014.

\bibitem{Cermak14mcheckSL}
P.~Cerm{\'{a}}k, A.~Lomuscio, F.~Mogavero, and A.~Murano.
\newblock {MCMAS-SLK}: A model checker for the verification of strategy logic
  specifications.
\newblock In {\em Proc. of {CAV}'14}, volume 8559 of {\em LNCS}, pages
  525--532. Springer, 2014.

\bibitem{Cermak15mcmas-sl-one-goal}
P.~Cerm{\'{a}}k, A.~Lomuscio, and A.~Murano.
\newblock Verifying and synthesising multi-agent systems against one-goal
  strategy logic specifications.
\newblock In {\em Proc. of {AAAI}'15}, pages 2038--2044, 2015.

\bibitem{Chatterjee10strategylogic}
K.~Chatterjee, T.A. Henzinger, and N.~Piterman.
\newblock Strategy {L}ogic.
\newblock {\em Inf. and Comp.}, 208(6):677--693, 2010.

\bibitem{Chen13prismgames}
T.~Chen, V.~Forejt, M.~Kwiatkowska, D.~Parker, and A.~Simaitis.
\newblock {PRISM-games}: A model checker for stochastic multi-player games.
\newblock In {\em Proc. of {TACAS}'13}, volume 7795 of {\em LNCS}, pages
  185--191. Springer, 2013.

\bibitem{Clarke89assGuar}
E.M. Clarke, D.E. Long, and K.L. McMillan.
\newblock Compositional model checking.
\newblock In {\em Proc. of {LICS}'89}, pages 353--362. {IEEE} Comput. Soc.
  Press, 1989.

\bibitem{Dima11undecidable}
C.~Dima and F.L. Tiplea.
\newblock Model-checking {ATL} under imperfect information and perfect recall
  semantics is undecidable.
\newblock {\em CoRR}, abs/1102.4225, 2011.

\bibitem{Huang14symbolic-epist}
X.~Huang and R.~van~der Meyden.
\newblock Symbolic model checking epistemic strategy logic.
\newblock In {\em Proc. of {AAAI}'14}, pages 1426--1432, 2014.

\bibitem{Jamroga20Pret-Uppaal}
W.~Jamroga, Y.~Kim, D.~Kurpiewski, and P.Y.A. Ryan.
\newblock Towards model checking of voting protocols in uppaal.
\newblock In {\em Proc. of {E-Vote-ID}'20}, volume 12455 of {\em LNCS}, pages
  129--146. Springer, 2020.

\bibitem{Jamroga19fixpApprox-aij}
W.~Jamroga, .~Knapik, D.~Kurpiewski, and {\L}.~Mikulski.
\newblock Approximate verification of strategic abilities under imperfect
  information.
\newblock {\em Artif. Int.}, 277, 2019.

\bibitem{Jamroga18Selene}
W.~Jamroga, M.~Knapik, and D.~Kurpiewski.
\newblock Model checking the {SELENE} e-voting protocol in multi-agent logics.
\newblock In {\em Proc. of {E-Vote-ID}'18}, volume 11143 of {\em LNCS}, pages
  100--116. Springer, 2018.

\bibitem{JPSDM20}
W.~Jamroga, W.~Penczek, T.~Sidoruk, P.~Dembinski, and A.W. Mazurkiewicz.
\newblock Towards partial order reductions for strategic ability.
\newblock {\em J. Artif. Intell. Res.}, 68:817--850, 2020.

\bibitem{Jamroga21natstrat-voting}
Wojciech Jamroga, Damian Kurpiewski, and Vadim Malvone.
\newblock Natural strategic abilities in voting protocols.
\newblock In {\em Proc. of {STAST}'20}, 2021.
\newblock To appear.

\bibitem{Kurpiewski19stv-demo}
D.~Kurpiewski, W.~Jamroga, and M.~Knapik.
\newblock {STV}: Model checking for strategies under imperfect information.
\newblock In {\em Proceedings of the 18th International Conference on
  Autonomous Agents and Multiagent Systems AAMAS 2019}, pages 2372--2374.
  IFAAMAS, 2019.

\bibitem{Kurpiewski21stv-demo}
D.~Kurpiewski, W.~Pazderski, W.~Jamroga, and Y.~Kim.
\newblock {STV+Reductions}: Towards practical verification of strategic ability
  using model reductions.
\newblock In {\em Proc. of {AAMAS}'21}, pages 1770--1772. ACM, 2021.

\bibitem{Lomuscio17mcmas}
A.~Lomuscio, H.~Qu, and F.~Raimondi.
\newblock {MCMAS}: An open-source model checker for the verification of
  multi-agent systems.
\newblock {\em Int. J. Soft. Tools Tech. Trans.}, 19(1):9--30, 2017.

\bibitem{Lomuscio10assGar}
A.~Lomuscio, B.~Strulo, N.G. Walker, and P.~Wu.
\newblock Assume-guarantee reasoning with local specifications.
\newblock In {\em Proc. of {ICFEM}'10}, volume 6447 of {\em LNCS}, pages
  204--219. Springer, 2010.

\bibitem{Lomuscio13assGar}
A.~Lomuscio, B.~Strulo, N.G. Walker, and P.~Wu.
\newblock Assume-guarantee reasoning with local specifications.
\newblock {\em Int. J. Found. Comput. Sci.}, 24(4):419--444, 2013.

\bibitem{Mik22assGua}
{\L}.~Mikulski, W.~Jamroga, and D.~Kurpiewski.
\newblock Towards assume-guarantee verification of strategic ability.
\newblock In {\em Proc. of {AAMAS}'22}.
\newblock to appear.

\bibitem{Mogavero14behavioral}
F.~Mogavero, A.~Murano, G.~Perelli, and M.Y. Vardi.
\newblock Reasoning about strategies: On the model-checking problem.
\newblock {\em ACM Trans. Comp. Log.}, 15(4):1--42, 2014.

\bibitem{Pnueli84assGuar}
A.~Pnueli.
\newblock In transition from global to modular temporal reasoning about
  programs.
\newblock In {\em Logics and Models of Concurrent Systems}, volume~13 of {\em
  {NATO} {ASI} Series}, pages 123--144. Springer, 1984.

\bibitem{Ryan10atemyvote}
P.Y.A. Ryan.
\newblock The computer ate my vote.
\newblock In {\em Formal Methods: State of the Art and New Directions}, pages
  147--184. Springer, 2010.

\bibitem{Schobbens04ATL}
P.Y. Schobbens.
\newblock Alternating-time logic with imperfect recall.
\newblock {\em Electr. Not. Theor. Comput. Sci.}, 85(2):82--93, 2004.

\bibitem{Tabatabaei16expressing}
M.~Tabatabaei, W.~Jamroga, and Peter Y.~A. Ryan.
\newblock Expressing receipt-freeness and coercion-resistance in logics of
  strategic ability: Preliminary attempt.
\newblock In {\em Proc. of {PrAISe@ECAI}'16}, pages 1:1--1:8. {ACM}, 2016.

\end{thebibliography}




\end{document}